\documentclass[prl,amsmath,amssymb,twocolumn]{revtex4}

\input epsf.tex
\usepackage{graphicx}
\usepackage{amsthm}

\newtheorem{lemma}{Lemma}

\newcommand{\be}{\begin{equation}}
\newcommand{\ee}{\end{equation}}
\newcommand{\ba}{\begin{eqnarray}}
\newcommand{\ea}{\end{eqnarray}}
\newcommand{\ban}{\begin{eqnarray*}}
\newcommand{\ean}{\end{eqnarray*}}

\newcommand*{\Hmin}{H_{\min}}
\newcommand*{\EC}{\mathrm{EC}}
\newcommand*{\leak}{\mathrm{leak}}
\newcommand*{\eps}{\varepsilon}
\newcommand*{\epsb}{\bar{\eps}}

\newcommand{\one}{\leavevmode\hbox{\small1\normalsize\kern-.33em1}}

\begin{document}

\title{Quantum cryptography with finite resources: unconditional
  security bound for discrete-variable protocols with one-way
  post-processing} \author{Valerio Scarani$^1$ and Renato Renner$^2$}
\affiliation{$^1$ Centre for Quantum Technologies and Department of Physics, National University of Singapore, 3 Science Drive 2, Singapore 117543, Singapore\\ $^2$ Institute for Theoretical Physics,
  ETH Zurich, 8093 Z\"urich, Switzerland} \date{\today}
\begin{abstract}
  We derive a bound for the security of QKD with finite resources
  under one-way post-processing, based on a definition of security
  that is composable and has an operational meaning. While our proof
  relies on the assumption of collective attacks, unconditional
  security follows immediately for standard protocols like
  Bennett-Brassard 1984 and six-states. For single-qubit
  implementations of such protocols, we find that the secret key rate
  becomes positive when at least $N\sim 10^5$ signals are exchanged
  and processed. For any other discrete-variable protocol,
  unconditional security can be obtained using the exponential de
  Finetti theorem, but the additional overhead leads to very
  pessimistic estimates.
\end{abstract}
\maketitle

\textit{Introduction.} Quantum cryptography, or more exactly quantum
key distribution (QKD), allows to distribute a secure key between two
authorized partners, Alice and Bob, connected by a quantum channel and
a public authenticated classical channel
\cite{review1,review2,review3}. First proposed in 1984 by Bennett and
Brassard (BB84, \cite{bb84}) and in 1991 by Ekert~\cite{Ekert91}, QKD
is the first offspring of quantum information science to reach the
level of applied physics and even commercial products. On the
theoretical side, much effort has been devoted to derive rigorous
bounds for security. However, almost all the available security bounds
hold true only if infinitely long keys are produced and processed. In
contrast, a practical QKD scheme can only use finite resources --- for
instance, Alice and Bob have limited computational power, and they can
only communicate a finite number of (qu)bits, resulting in keys of
finite length.

The security of finite-length keys has been studied first
in~\cite{ina07} and later in~\cite{wat,hay} for the BB84 protocol, as
well as in~\cite{mey06} for a larger class of protocols. The
applicability of these results is, however, limited: Ref.~\cite{mey06}
considers only a restricted class of attacks; in
Refs~\cite{ina07,wat,hay}, the underlying notion of security is
\emph{not composable}~\cite{KRBM07}, which means that the generated
keys are not secure enough to be used in applications, e.g., for
encryption (more below). A more recent work~\cite{hay2}, which focuses
on a practical implementation of BB84 and has already been used in an
experiment~\cite{hase}, uses a definition of security which is
probably composable, although the issue is not discussed. In this
Letter, we provide a security bound for discrete-variable QKD
protocols with finite resources and with respect to a
\emph{composable} security definition, based on the formalism
developed by one of us \cite{rennerthesis}. As first case studies, we
apply it to BB84 and to the six-states protocol~\cite{bru98,bec99}
when implemented with single qubits.

\textit{Definition of security.} In the existing literature on QKD,
not only the analysis, but also the very \emph{definition} of security
is mostly limited to the asymptotic case; and we therefore need to
revisit it here. Most generally, the security of a key $K$ can be
parametrized by its \emph{deviation} $\eps$ from a \emph{perfect key},
which is defined as a uniformly distributed bit string whose value is
completely independent of the adversary's knowledge.  In an
\emph{asymptotic} scenario, a key $K$ of length $\ell$ is commonly
said to be \emph{secure} if this deviation $\eps$ tends to zero as
$\ell$ increases.  In the \emph{non-asymptotic} scenario studied here,
however, the deviation $\eps$ is always finite.  This makes it
necessary to attribute an \emph{operational interpretation} to the
parameter $\eps$. Only then it is possible to choose a meaningful
security threshold (i.e., an upper bound for $\eps$) reflecting the
level of security we are aiming at. %% ~\footnote{This is less relevant in
%%   the asymptotic case, because most reasonable
%%    measures for the deviation $\eps$ lead to asymptotically equivalent definitions.}.
Another practically relevant requirement that we need to take into
account is \emph{composability} of the security definition.
Composability guarantees that a key generated by a QKD protocol can
safely be used in applications, e.g., as a one-time-pad for message
encryption. Although this requirement is obviously crucial for
practice, it is not met by most security definitions considered in the
literature~\cite{KRBM07}.

In contrast to that, the results derived in this Letter are formulated
in terms of a security definition that meets both requirements, i.e.,
it is composable and, in addition, the parameter $\eps$ has an
operational interpretation. The definition we use was proposed
in~\cite{RenKoe05,BHLMO05}: for any $\eps \geq 0$, a key $K$ is said
to be \emph{$\eps$-secure with respect to an adversary $E$} if the
joint state $\rho_{K E}$ satisfies \ba \frac{1}{2} \bigl\| \rho_{K E}
- \tau_K \otimes \rho_E \bigr\|_1 &\leq& \eps \ , \ea where $\tau_K$
is the completely mixed state on $K$. The parameter $\eps$ can be seen
as the maximum probability that $K$ differs from a perfect key (i.e.,
a fully random bit string)~\cite{RenKoe05}.  Equivalently, $\eps$ can
be interpreted as the \emph{maximum failure probability}, where
failure means that ``something went wrong'', e.g., that an adversary
might have gained some information on $K$.  From this perspective, it
is also easy to understand why the definition is composable. In fact,
the failure probability of any cryptosystem that uses a perfect secret
key only increases by (at most) $\eps$ if the perfect key is replaced
by an $\eps$-secure key. In particular, because one-time pad
encryption with a perfect key has failure probability $0$ (the
ciphertext gives zero information about the message), it follows that
one-time-pad encryption based on an $\eps$-secure key remains
perfectly confidential, except with probability at most~$\eps$.

% **** Here, one might directly start with the entanglement-based
% picture. The states $\ket{a}$ and $\ket{b}$ are not needed in the
% seque. ****
\textit{Protocol.} A QKD protocol starts with the distribution of
quantum signals. In this Letter, we take an \emph{entanglement-based
  view}, that is, after this distribution step, Alice and Bob share
$N$ (entangled) particle pairs, whose joint state we denote by
$\rho_{A^N B^N}$. Next, Alice and Bob apply individual measurements to
their particles to get classical data. For definiteness, we focus on
protocols that use two-dimensional quantum systems (qubits) and von
Neumann measurements, resulting in $N$ correlated pairs of bits. Then,
in a \emph{parameter estimation} step, Alice and Bob reveal a random
sample consisting of $m$ of these pairs (using a public communication
channel) which allows them to estimate the \emph{statistics}
$\lambda_{(a,b)}$ of their data, i.e., the relative frequency of the
symbols. The protocol may also specify a \emph{sifting} phase, in
which some items are discarded.

%: in
%the BB84 and six-states protocols, for instance, Alice and Bob keep
%only the items that have been generated by measurements in coincident
%bases, i.e., $A_i=B_i$.

%% Alice and Bob exchange $N$ quantum signals; $N_s$ symbols remain after
%% sifting, $m$ among which are used for parameter estimation. The
%% remaining $n=N_s-m$ symbols define the raw key (actually, in full
%% generality, Alice and Bob may find it advantageous to apply to these
%% symbols some, possibly blockwise, pre-processing, whose outcome is the
%% raw key; but we neglect this possibility here).  Note that $n$ and $m$
%% are not fixed parameters: for any given $N$, Alice and Bob have to
%% make the choice that optimizes $K$, i.e. find the best compromise
%% between a long raw key and an accurate parameter estimation. In the
%% asymptotic limit $N\rightarrow\infty$, an arbitrarily small fraction
%% of signals provides a sufficiently accurate parameter estimation, so
%% basically all the signals can be used for the key: $n/N\rightarrow 1$
%% and $m/N\rightarrow 0$.

At this stage, both Alice and Bob hold a string of $n \leq N-m$ bits,
called \emph{raw key}, denoted by $X^n$ and $Y^n$, respectively.
These raw keys are generally only partially correlated and only
partially secret.  But|and this is where quantum physics plays a
role|the maximum information that an eavesdropper Eve might have
gained during the protocol, in the following denoted $E^n$, can be
computed solely from the statistics $\lambda_{(a,b)}$.  This allows
Alice and Bob to transform the raw key pair into a fully secure key
$K$ of length $\ell \leq n$, using some purely classical procedure, in
the following called \emph{post-processing}.  In this Letter, we focus
on \emph{one-way post-processing} consisting of two steps, called
\emph{error correction} (also known as \emph{information
  reconciliation}) and \emph{privacy amplification}.  For the error
correction, Alice sends some information on her raw key $X^n$ over the
public channel, allowing Bob, who already knows $Y^n$, to compute a
guess for $X^n$.
%% ~\footnote{The
%%   roles of Alice and Bob can of course be interchanged. This is
%%   sometimes referred to as \emph{reverse reconciliation}}
Finally, privacy amplification is applied to turn $X^n$ into a fully
secure key $K$. This is typically done by \emph{two-universal
  hashing}~\footnote{J.~L.  Carter, M.~N.
  Wegman, Journal of Computer and System Sciences \textbf{18}, 143
  (1979); M.~N. Wegman, J.~L. Carter, \textit{idem} \textbf{22}, 265
  (1981)}.

\textit{Asymptotic analysis.} The one-way protocol described above has
been studied extensively over the past few years, mostly in an
asymptotic scenario where the size of the raw key tends to
infinity. In this case, a commonly used figure of merit is the
\emph{sifted key rate} $r'$, defined as the ratio $r' := \lim_{n \to
  \infty} \frac{\ell(n)}{n}$ between the number $\ell(n)$ of generated
key bits and the size $n$ of the raw key. Devetak and
Winter~\cite{DevWin05} have proved that, under the assumption of
collective attacks (see below),
\begin{align}
  r' & = H(X|E) - H(X|Y) \ , \label{eq:entropyrateasym}
\end{align}
where $H(.|.)$ is the conditional von Neumann entropy, evaluated after
the sifting step|note that, when both systems are classical as in
$H(X|Y)$, von Neumann entropy becomes Shannon entropy.  The expression
says that the sifted key rate $r'$ is equal to the uncertainty that
Eve has on the raw key bits $X$, minus Bob's uncertainty: a very
intuitive statement after all. Multiplying the sifted key rate $r'$
with the ratio $\frac{n}{N}$ of raw key bits per signal gives the
\emph{key rate per signal} $r$, which is an indicator for the
asymptotic performance of the overall protocol. For many schemes, the
ratio $\frac{n}{N}$ can be chosen arbitrarily close to one for
sufficiently large $N$, because a small fraction $m<<N$ of signals
provides a sufficiently accurate parameter estimation; in this case,
the key rate per signal $r$ and the sifted key rate $r'$ are
asymptotically equal.

\textit{Non-asymptotic analysis.} When the number $N$ of exchanged
quantum signals is finite, the above considerations are no longer
sufficient. For example, since $n + m \leq N$, one has to find a
trade-off between the length of the raw key $n$ and the precision of
parameter estimation, which depends on the sample size
$m$. \textit{Imperfect parameter estimation} is however not the only
deviation from the asymptotic case. The \textit{performance of an
  error correction procedure} $\EC$ might --- and actually does in
practical realizations --- perform worse than the theoretical
limit. For our security analysis, the main characteristics of $\EC$
are the number of bits that need to be transmitted over the public
channel (carrying information on $X^n$), in the following denoted
$\leak_{\EC}$, and the error probability $\eps_{\EC}$, i.e., the
probability that Bob computes a wrong guess for $X^n$.  Finally, as
discussed above, \textit{the security of a key generated from finite
  resources is always finite}: the length of the extractable secret
key depends on the desired security $\eps$ of the final key.

Our goal is to find the generalization of~\eqref{eq:entropyrateasym}
for QKD with finite resources, and to use it to compute $r$ for given
$(N,\eps,\leak_{\EC},\eps_{\EC})$ after optimizing over the choices of
other possible parameters. The analysis will be based on the tools
developed in~\cite{rennerthesis}. It particular, it relies on a
generalization of the von Neumann entropy~ \footnote{The conditional
  von Neumann entropy evaluated for a density operator $\sigma_{A B}$
  can be expressed asymptotically in terms of the smooth min-entropy
  evaluated for i.i.d.\ states $\rho_{A^n B^n} = \sigma_{A B}^{\otimes
    N}$, i.e., $H(A|B)_{\sigma_{A B}} = \lim_{\varepsilon \to 0}
  \lim_{N \to \infty} \frac{1}{n} \Hmin^{\varepsilon}(A^n|B^n)$.},
called \emph{smooth min-entropy}. For any bipartite density operator
$\rho_{A B}$ and $\varepsilon \geq 0$, the smooth min-entropy
$\Hmin^\varepsilon(A|B)$
% ~\footnote{The smooth min-entropy is sometimes denoted
%   $H_{\infty}^\varepsilon(A|B)$.}
is defined as the maximum,
taken over all density operators $\bar{\rho}_{A B}$ that are
$\eps$-close to $\rho_{A B}$, of the quantity
\begin{align*}
  \Hmin(A|B) := -\log_2 \min \{\lambda \! > \! 0 \!: 
  \exists \sigma_B \! : \bar{\rho}_{A B} \leq \lambda \, \mathrm{id}_A \! \otimes \!
  \sigma_B \}
\end{align*}
where $\mathrm{id}_A$ denotes the identity operator on subspace $A$
and $\sigma_B$ is any density operator on subspace $B$. The
significance of the smooth min-entropy stems from the fact that it
characterizes the number of uniform bits that can be extracted by
privacy amplification.

As a starting point, a formula for the number of final key bits $\ell$
can be obtained as a straightforward generalization of Lemma~6.4.1
in~\cite{rennerthesis}:
\begin{lemma} \label{lem:keylength} The key agreement protocol
  described above generates an $\eps$-secure key if, for some
  $\bar{\eps} \geq 0$,
\begin{align} \label{eq:keylength}
    \ell \leq \Hmin^{\epsb}(X^n|E^n) - \leak_{\EC} - 2 \log_2 {\textstyle \frac{1}{2 (\eps - \bar{\eps} - \eps_{\EC} )}} \ .
\end{align}
\end{lemma}

Lemma~\ref{lem:keylength} shows explicitly the two-step nature of
one-way post-processing: for error correction, Alice has to send a bit
string $C$ of length $\leak_{\EC}$ to Bob over the public channel,
hence, reducing Eve's uncertainty by the same amount. Privacy
amplification then extracts a key whose length roughly corresponds to
Eve's uncertainty after error correction, which is given by
$\Hmin^{\epsb}(X^n| C E^n) \geq \Hmin^{\epsb}(X^n|E^n) -
\leak_{\EC}$~\footnote{If privacy amplification is applied to
  individual blocks rather than to the overall raw key, then
  expression~\eqref{eq:keylength} needs to be evaluated for each of
  the blocks separately.}.

To go further, we have to evaluate the smooth min-entropy
$\Hmin^{\epsb}(X^n|E^n)$. This evaluation is easy in the case of
collective attacks, i.e., under the assumption that Alice and Bob (in
an entanglement-based view) initially share a state of the form
$\rho_{A^N B^N}=(\sigma_{\bar{A} \bar{B}})^{\otimes N}$ with
$\sigma_{\bar{A} \bar{B}}$ a two-qubit state. Indeed, in this case one
can also assume $\rho_{X^n E^n}=(\sigma_{\bar{X} \bar{E}})^{\otimes
  n}$ without loss of generality, since all purifications of $\rho_{A
  B}$ are equivalent under a local unitary operation by Eve, and there
exists clearly a purification with that property.  However, the
statistics $\lambda_{(a,b)}$ acquired during parameter estimation
generally only gives a partial characterization of $\sigma_{\bar{X}
  \bar{E}}$.  Lemma~\ref{lem:asym} below
\footnote{Lemma~\ref{lem:asym} is a Corollary~3.3.7
  of~\cite{rennerthesis}; we correct a typo (the ``1'' under the
  square root must be also divided by $n$) and set $d = 2$ as we are
  assuming that $X^n$ is a bit string.} provides a lower bound on
$\Hmin^{\epsb}(X^n|E^n)$, given that $\sigma_{\bar{X}\bar{E}}$ is
contained in a set $\Gamma$ compatible with the statistics
$\lambda_{(a,b)}$, except with probability~$\epsb'$.

\begin{lemma} \label{lem:asym} For any $\epsb > \epsb'$, the smooth
  min-entropy of the state $\rho_{X^n E^n}$ described above is lower
  bounded by
  \begin{equation} \label{eq:Hminbound}
    \Hmin^{\epsb}(X^n | E^n)
  \geq
    n \bigl( \min_{\sigma_{\bar{X} \bar{E}} \in \Gamma} H(\bar{X} | \bar{E}) - \delta \bigr)
  \end{equation}
  where $\delta := 7 \sqrt{\frac{\log_2(2/(\epsb -
      \epsb'))}{n}}$.
\end{lemma}

The description of the set of states $\Gamma$ takes into account the fact that
the parameter estimation has been made on a sample of finite size $m$.
A quantitative version of the law of large numbers (see e.g.~ Theorem 12.2.1 and Lemma
12.6.1 in \cite{cover}) yields the following statement:%%  if we have $M$
%% realizations of a random variable that can take $d$ different values,
%% then $\mbox{Prob}\left(\|\lambda_M-\lambda_{\infty}\|\geq
%%   \xi\right)\equiv \epsb\,'\leq (M+1)^de^{-\demi M\xi^2}$, where
%% $\lambda_{\infty}$ are the ideal statistics. Therefore we have

\begin{lemma} \label{lem:estimate} If the statistics $\lambda_m$ are
  obtained by measurements of $m$ samples of $\sigma$ according to a
  POVM with $d$ outcomes then, for any $\epsb\,'>0$,  $\sigma$ is
  contained in the set
\[
  \Gamma_{\xi}=\left\{\sigma:\,\|\lambda_m-\lambda_{\infty}(\sigma)\|\leq\xi := {\textstyle \sqrt{ \frac{2
    \ln(1/\epsb\,') + d\, \ln(m+1)}{m}}}\right\} 
\]
except with probability $\epsb\,'$, where $\lambda_{\infty}(\sigma)$
denotes the probability distribution defined by the
POVM applied to $\sigma$.
\end{lemma}

The three Lemmas together yield the desired generalization
of~\eqref{eq:entropyrateasym}: \ba r' &=&
H_\xi(X|E)-\big(\leak_{\EC}+\Delta\big)/n%\frac{\leak_{\EC}}{n}-\frac{\Delta}{n}
\label{raten}\ea with
$H_\xi(X|E)=\min_{\sigma_{\bar{X} \bar{E}} \in \Gamma_{\xi}} H(\bar{X}
| \bar{E})$ and $\Delta = 2 \log_2 {1/[2 (\eps - \epsb - \eps_{\EC})]}
+ 7 \sqrt{n\log_2(2/(\epsb - \epsb\,'))}$. We recall that
$(N,\eps,\leak_{\EC},\eps_{\EC})$ are parameters of the protocol
implementation, while $n$, $m$, $\epsb$ and $\epsb\,'$ must be chosen
as to maximize $r=(n/N) r'$ under the constraints $n+m\leq N$ and
$\eps - \eps_{\EC}>\epsb>\epsb\,'\geq 0$.

In general, \eqref{raten} is valid only for collective attacks because
of the estimate \eqref{eq:Hminbound} of $\Hmin^{\epsb}(X^n
|E^n)$. However, it has been proved that the assumption of collective
attacks can be made without loss of generality for the BB84 and the
six-states protocols~\cite{gotlo,KGR} (see open issues for the
discussion of a more general approach based on the exponential de
Finetti theorem~\cite{rennerthesis,symindep}). To illustrate the
bound~\eqref{raten}, we move on to derive the explicit expressions of
$H_\xi(X|E)$.

\textit{BB84.} We consider an asymmetric version of BB84 \cite{lochau}: the key is
obtained from measurements in one basis ${\cal B}_0$ chosen both by
Alice and Bob with probability $p_0$; the complementary basis ${\cal
  B}_1$, chosen with probability $p_1=1-p_0$ is used for parameter
estimation. So $n=Np_0^{\,2}$ and $m=Np_1^{\,2}$, while $2Np_0p_1$ signals are discarded in sifting. 
The computation of $H_\xi(X|E)$ can be done in full along the usual lines, see e.g. Appendix A of \cite{review3}. More directly, notice that, in this term, the only finite-key effect is the imperfection of the statistics. Knowing the asymptotic value $H(\bar{X} |
\bar{E})=1-h(e_1)$ where $e_1$ is the error rate in the basis ${\cal B}_1$ (\textit{phase error}), it is obvious that the worst-case estimate of $\lambda_{(a,b)}\equiv e_1$ is $\tilde{e}_1=e_1+\xi(m,d=2)$ because the POVM has two outcomes (same vs different bits). Therefore
\ba H_\xi(X|E)&=& 1-h(\tilde{e}_1)\,.
\ea

\textit{Six-states.} We consider an asymmetric version of the six-states protocol: the key is
obtained from measurements in one basis ${\cal B}_0$ chosen both by
Alice and Bob with probability $p_0$; the complementary bases ${\cal B}_1$ and ${\cal B}_2$, chosen with equal probability $q=\frac{1-p_0}{2}$, are used for parameter estimation. Sifting yields $n=Np_0^{\,2}$ and $m_1=m_2=Nq^{\,2}$ while the remaining signals are discarded. Similarly as above, the asymptotic formula (for $e_1=e_2$, a case that minimizes it) can be immediately translated into
\ba H_\xi(X|E)&=& (1-\tilde{e}_0)\left[1-h\left(\frac{1-\tilde{e}_1-\tilde{e}_0/2}{1-\tilde{e}_0}\right)\right]
\ea
with $\tilde{e}_1=e_1+\xi(m_1,d=2)$ and $\tilde{e}_0=e_0+\xi(n,d=2)$, because $e_0$ is estimated on the $n$ bits of the raw key.

\begin{figure}
\includegraphics[scale=0.6]{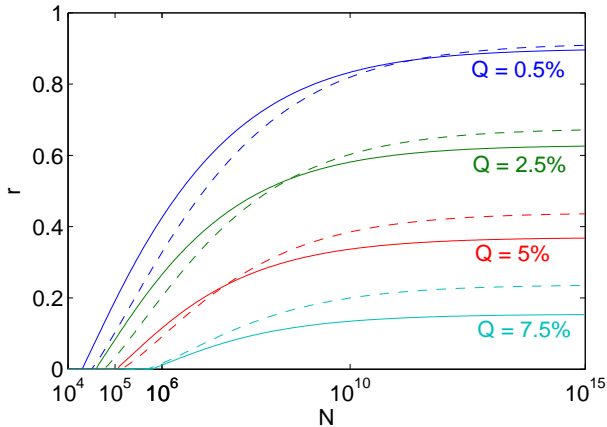}
\caption{(color online) Lower bound for the key rate $r$ as a function of the number of exchanged quantum signals $N$, for the BB84 (full lines) and the six-states protocol (dashed lines); values: $\eps=10^{-5}$, $\eps_{EC}=10^{-10}$, $\leak_{\EC}/n=1.2h(Q)$, and several $Q=e_0=e_1$.} \label{figfinite}
\end{figure}

\textit{Plots.} For an \textit{a priori} estimate of our bounds, we have supposed as usual that parameter estimation yields $e_0=e_1\equiv Q$; imperfect EC has been characterized by $\leak_{\EC}/n=1.2h(Q)$ and $\eps_{EC}=10^{-10}$ based on the performances of real codes \footnote{Communications by O.Gay (idQuantique, Geneva), M. Peev and C. Pacher (ARC Seibersdorf, Vienna) and C. Kurtsiefer (NU Singapore).}. The optimization was done numerically; in particular, the optimal value of $p_1$ was found to be approximately $\frac{1}{n_b}(N/N_0)^{-1/4}$, $N_0$ being the smallest $N$ such that $r>0$ and $n_b=2$ for BB84 and 3 for six-states. The results are shown in Fig.~\ref{figfinite}. The slight difference between the two protocols is due to the fact that six-states estimates more parameters than BB84: the rates are in principle higher because the bound on Eve's information is tighter, but, for short keys, more signals must be devoted to the estimation. These plots do not depend very critically on the value $\eps$; in particular, even for $\eps\geq 10^{-2}$ our bounds are tighter than those computed in \cite{mey06} for a limited class of attacks on the six-states protocol.

\textit{Open issues.---} We point out two directions for future
work. \textit{First:} The results we have presented here are not
necessarily tight: better estimates might lead to more optimistic
bounds on the security. Lemmas~\ref{lem:keylength}--\ref{lem:estimate}
can be shown to be optimal up to an additive term of the order $\log
1/\varepsilon$. So basically there is room for improvement only in the
performance of error correction schemes. \textit{Second:}
Formula~\eqref{raten} has been derived under the assumption of
collective attacks and provides full security for the BB84 and the
six-states protocols only thanks to specific
symmetries~\cite{gotlo,KGR}. To get a fully general statement, one
might invoke a quantum version of de Finetti's representation theorem
as proposed in~\cite{symindep}, which, in the asymptotic case, implies
that security against general attacks follows from security against
collective attacks. This technique, however, gives rise to additional
deviations (see Theorem 6.5.1 of~\cite{rennerthesis} for explicit
formulae) which are significant in a non-asymptotic scenario and lead
to very pessimistic bounds. To improve them, a tighter variant of de
Finetti's theorem, or some new ideas, might be required.

\textit{Acknowledgments.}--- We thank J.-C. Boileau, M. Hayashi,
N. L\"utkenhaus and other participants to the workshop ``Tropical
QKD'' (Waterloo, Canada, June 2007) for clarifying discussions. This
work is supported by the National Research Foundation and Ministry of
Education, Singapore, by HP Labs Bristol, and by the European Union
through the projects SECOQC and SCALA.

\end{document}